\documentclass[twocolumn,showpacs,preprintnumbers,amsmath,amssymb]{revtex4-1} \usepackage{dcolumn}

\usepackage{graphicx} \usepackage{bm}

\usepackage{graphicx} \usepackage{bm} \usepackage{dcolumn}

\newcommand{\hH}{\hat{\cal H}}
\newcommand{\hZ}{\hat{\cal Z}}
\newcommand{\hU}{\hat{\cal U}}
\newcommand{\hS}{\hat{\cal S}}
\newcommand{\hB}{\hat{\beta}}
\newcommand{\hl}{\hat{\lambda}}

\begin{document}
\title{Transformation of bound states of relativistic hydrogen-like atom into two-component form}
\date{\today}
\author{Tomasz M. Rusin}
\address{Orange Customer Service sp. z o. o., Al. Jerozolimskie, 02-326 Warsaw, Poland}

\email{Tomasz.Rusin@orange.com}


\begin{abstract}
A single-step Eriksen transformation of~$1S_{1/2}$,~$2P_{1/2}$ and~$2P_{3/2}$
states of the relativistic hydrogen-like atom is performed exactly by expressing each transformed
function (TF) as a linear combination of eigenstates of the Dirac Hamiltonian.
The transformed functions, which are four-component spinors with vanishing two lower components,
are calculated numerically and have the same symmetries as
the initial states. For all nuclear charges~$Z \in [1\ldots 92]$ a contribution of the initial
state to TFs exceeds 86\% of the total probability density.
Next large contribution to TFs comes from continuum states
with negative energies close to~$-m_0c^2-E_b$, where~$E_b$ is the binding energy
of initial state. Contribution of other states to TFs is less
than~$0.1\%$ of the total probability density. Other components of TFs
are nearly zero which confirms both validity of the Eriksen transformation and
accuracy of the numerical calculations. The TFs of~$1S_{1/2}$ and~$2P_{1/2}$ states are close
to~$1s$ and~$2p$ states of the nonrelativistic hydrogen-like atom, respectively, but the TF
of~$2P_{3/2}$ state differs qualitatively from the~$2p$ state. Functions
calculated with use of a linearized Eriksen transformation, being equivalent to the second order
Foldy-Wouthuysen transformation, are compared
with corresponding functions obtained by Eriksen transformation.
A very good agreement between both results is obtained.
\end{abstract}

\maketitle

\section{Introduction} \label{Sec_Intro}

In their pioneering work, Foldy and Wouthuysen~(FW) introduced the method of
separating positive and negative energy states of the Dirac Hamiltonian~\cite{Foldy1950}.
In the presence of external fields the odd terms of the
lowest order in~$1/(m_0c^2)$ are removed to the
desired accuracy by a sequence of unitary transformations.
As a result, the states of the Dirac Hamiltonian having positive energies are transformed
to the two-component form, in which two upper components are nonzero
and two lower components vanish, up to expected accuracy. This way
one reduces the Dirac equation to two equations for two-component spinors
describing states of positive and negative energies, respectively.
The widely used second-order FW-transformation converts
the Dirac Hamiltonian into a Schrodinger Hamiltonian with
relativistic corrections:~$\hat{p}^4$ term, the spin-orbit interaction and the Darwin term.
Works considering higher-order FW transformations are discussed in Ref.~\cite{deVries1970}.

Another possibility to transform the Dirac Hamiltonian to a two-component form
was proposed by Douglas and Kroll~(DK), Ref.~\cite{Douglas1974}.
In this approach, used mostly in the quantum chemistry, one performs a series of transformations leading
to an expansion of the Dirac Hamiltonian in orders of external potential, see~\cite{Douglas1974,Jansen1989}.
In each step of DK transformation the odd terms of the lowest order are removed up to expected accuracy.
Using this method it is possible to include many-electron effects into real
atomic systems~\cite{Nakajima2003}.
For a review of works related to this subject see Ref.~\cite{Reiher2015}.

Practical limitations in applying higher-order FW or DK transformations are due to
complicated calculations since one has to manipulate increasing numbers
of noncommuting operators. To overcome these limitations several approaches were proposed,
removing odd terms to a sufficiently high order with the use of numerical
methods~\cite{Silenko2003,Reiher2004}. The resulting~$2\times 2$ Schrodinger-like equations
with relativistic corrections are then solved numerically
for energies and wave functions. A good agreement between approximate and exact Dirac energies
was reported~\cite{Leeuwen1994,Lenthe1996,Barysz2002,Reiher2004}.

One can avoid complicated calculations in special cases using transformations
separating exactly the Dirac Hamiltonian into a two-component form.
Such a transformation for free relativistic electrons was proposed by
Foldy and Wouthuysen~\cite{Foldy1950}. Case~\cite{Case1954} found an exact form of
the FW transformation for the presence of a constant magnetic field.
Tsai~\cite{Tsai1973} and Weaver~\cite{Weaver1975} reported exact FW transformation
for the presence of a magnetic field and electro-weak interactions.
Moss and Okninski~\cite{Moss1976} pointed out that there exist
several transformations separating positive and negative states of the Dirac Hamiltonian,
leading to similar but not identical results for transformed functions.
Nikitin~\cite{Nikitin1998} reported on FW-like transformations for a constant electric field,
dipole potential and some classes of external fields with special symmetries.
The common weakness of the above methods is inability to generalize
their results to arbitrary potentials.

There exists in the literature several examples of functions transformed with the use of
FW-like or DK-like transformations.
A calculation of transformed functions and the transformation kernel for the FW transformations
for free Dirac electrons were given by Rusin and Zawadzki in Ref.~\cite{Rusin2011}, and for the presence
of a magnetic field in Ref.~\cite{Rusin2012}.
In the latter paper, the analytical expression for a transformed Gaussian wave packet was obtained.
Neznamov and Silenko~\cite{Silenko2008,Neznamov2009}
analyzed properties of functions transformed with the use of the FW-like transformations
and showed that the lower components of resulting functions are in the second order of~$1/(m_0c^2)$.
In several works related to DK-like transformations the resulting functions were
obtained numerically~\cite{Leeuwen1994,Lenthe1996,Barysz2002,Reiher2004,Reiher2015}.

A transformation of the Dirac Hamiltonian for any potential converting
it into a block-diagonal form was proposed by Eriksen in Refs.~\cite{Eriksen1958,Eriksen1960}.
This transformation is performed in a single step by an unitary operator~$\hU$
being a nonlinear function of the Dirac Hamiltonian. Because of its nonlinearity,~$\hU$
was usually approximated by a finite series in powers of~$1/(m_0c^2)$~\cite{Eriksen1958,deVries1970}.
The validity of Eriksen transformation and its power-series expansion was confirmed by
de Vries~\cite{deVries1970} and Silenko~\cite{Silenko2013}.
In the lowest-order terms, the Eriksen transformation agrees with the results obtained
by the FW method, but higher order terms differ~\cite{deVries1970,Neznamov2009}.

The subject of the present work is to transform bound states of the Dirac Hamiltonian
with a nontrivial potential by the Eriksen operator~$\hU$ without expanding~$\hU$ in a power series.
In our approach we concentrate on the transformed functions and their properties and not on the
transformed operators. Our calculations are performed for the relativistic hydrogen-like atom
whose spectrum consists of both bound and continuum states.
To illustrate our method and results the transformation is performed numerically for
the three lowest bound states of the relativistic hydrogen-like atom:~$1S_{1/2}$,~$2P_{1/2}$ and~$2P_{3/2}$,
for several values of the nuclear charge~$Z$.
To our knowledge there was no attempt to calculate functions transformed by the
single-step Eriksen transformation for the Dirac Hamiltonian whose eigenfunctions are bound states.

It should be reminded that solving the Dirac equation (analytically or numerically) one directly obtains
all its eigenvalues and eigenstates that can be used for calculations of observables, and no
further transformation is needed. The Dirac equation transformed to the block-diagonal form
can also be used for calculation of the observables, and both approaches yield the same results
because the wave functions in both representations are related each to other
by the unitary transformation. Thus, the choice of representations of the Dirac equation
depends on its convenience in further calculations or applications.

The paper is organized as follows. In Sec.~\ref{Sec_EC} we describe our approach,
in Sec.~\ref{Sec_SE} specify wave functions of the relativistic hydrogen-like atom.
In Sec.~\ref{Sec_Res} we show results of calculations and in Sec.~\ref{Sec_Disc} we discuss our
results. The paper is concluded by a Summary.
In two Appendices we specify details of our calculations.

\section{Eriksen transformation} \label{Sec_EC}

Let us consider the Dirac Hamiltonian describing a relativistic electron in the presence of
the Coulomb potential created by the atomic nucleus
\begin{equation} \label{ET_hH}
\hH = c\sum_{i=x,y,z}{\bf \hat{\alpha}}_i {\bm \hat{p}}_i + \hB m_0c^2 - \frac{Ze^2}{4\pi\epsilon_0 r},
\end{equation}
in which~$\hat{\alpha}_i$ and~$\hB$ are Dirac matrices in the standard
notation,~$|e|$ and~$m_0$ are the electron charge and mass, respectively,
and~$Z \in [1\ldots 92]$ is the nuclear charge.
Both eigenenergies and eigenstates of~$\hH$ are known analytically. The latter are given by
four-component spinors. The spectrum of~$\hH$ consists of an infinite set of bound states
having positive energies below~$E=+m_0c^2$, and two sets of continuum states
having energies above~$+m_0c^2$ and below~$-m_0c^2$, respectively.

The Eriksen transformation is defined by the following unitary operator~\cite{Eriksen1958}
\begin{equation} \label{ET_hU}
 \hU = \frac{1}{\sqrt{2+\hB\hl + \hl\hB}}\left(1+\hB\hl\right) \equiv \hZ + \hZ\hB\hl.
\end{equation}
The above equation defines the operator~$\hZ$
\begin{equation} \label{ET_hZ}
 \hZ=\frac{1}{\sqrt{2+\hB\hl + \hl\hB}}.
 \end{equation}
The sign operator~$\hl$ is defined as
\begin{equation} \label{ET_hT}
 \hl = \frac{\hH}{\sqrt{\hH^2}}.
\end{equation}
It is assumed that~$E=0$ is not an eigenenergy of~$\hH$, which holds for~$\hH$ in Eq.~(\ref{ET_hH})

Let~$\langle {\bm r}|{\rm n\rangle} = (\psi_1,\psi_2,\psi_3,\psi_4)^T$ be an arbitrary eigenfunction
of~$\hH$ corresponding to a positive energy. After the Eriksen transformation there is (see Appendix~\ref{Appendix_Z})
\begin{equation} \label{ET_Un}
\hU \left(\begin{array}{c}     \psi_1 \\     \psi_2 \\ \psi_3 \\ \psi_4 \end{array} \right) =
    \left(\begin{array}{c} \hU \psi_1 \\ \hU \psi_2 \\      0 \\      0 \end{array} \right),
\end{equation}
i.e., the transformed function has vanishing two lower components.
Because~$\hU$ in Eq.~(\ref{ET_hU}) is a non-linear function of the
differential operator~$\hat{\rm \nabla}$ and the position operator~$\hat{\bm r}$, it is not possible to
express~$\hU \psi_1$ and~$\hU \psi_2$ in a closed form.

To overcome this problem one notes that for an arbitrary eigenstate~$|{\rm n}\rangle$ of~$\hH$
having energy~$E_{\rm n} \neq 0$ there is
\begin{equation} \label{ET_Theta}
 \hl|{\rm n}\rangle = \frac{E_{\rm n}}{|E_{\rm n}|}|{\rm n}\rangle \equiv \epsilon_{\rm n} |{\rm n}\rangle,
\end{equation}
where~$\epsilon_{\rm n}=\pm 1$ is the sign of~$E_{\rm n}$.
Thus, using Eq.~(\ref{ET_Theta}) one can calculate the matrix elements of the
numerator of~$\hU$ in Eq.~(\ref{ET_hU}) between the eigenstates of~$\hH$ in Eq.~(\ref{ET_hH}). Similarly,
expanding~$\hZ$ in Eq.~(\ref{ET_hZ}) in a power series of~$(\hB\hl+\hl\hB)$ one can calculate matrix
elements of~$\hZ$ between the eigenstates of~$\hH$.
Thus, in spite of the fact that we do not know explicit form of~$\hU$ in Eq.~(\ref{ET_hU}),
we have a way to calculate its {\it matrix elements} between all eigenstates of~$\hH$.
This observation indicates the method of calculating the transformed functions in Eq.~(\ref{ET_Un}).

Let~$|{\rm n}_0\rangle$ and~$|{\rm \chi}\rangle =\hU|{\rm n}_0\rangle$ be an eigenstate of~$\hH$
and its Eriksen-transformed counterpart, respectively. Since~$|{\rm n}\rangle$
form a complete set of states in the space of four-component spinors, we may express~$|{\rm \chi}\rangle$
as a linear combination of~$|{\rm n}\rangle$
\begin{equation} \label{ET_chi}
 |{\rm \chi}\rangle = \sum_{\rm n} a_{\rm n} |{\rm n}\rangle, \hspace*{1em} a_{\rm n}=\langle {\rm n}|\hU|{\rm n}_0\rangle.
\end{equation}
Equation~(\ref{ET_chi}) indicates the way to calculate the function~$|{\rm \chi}\rangle$. First, one
selects large but finite set of eigenstates of~$\hH$
in Eq.~(\ref{ET_hH}), including both bound and continuum states having positive or negative energies.
Then we calculate (analytically or numerically) the matrix
elements~$\langle {\rm n}|\hU|{\rm n}_0\rangle$ between
all states in the selected set. Finally, we calculate~$\langle{\bm r}|{\rm \chi}\rangle$ as a sum
over all eigenfunctions~$\langle {\bm r}|{\rm n}\rangle$ of~$\hH$, as given in Eq.~(\ref{ET_chi}).
In our approach we make only one approximation, namely we truncate the infinite set of eigenstates
of~$\hH$ into a finite one.

Below we describe consecutive steps necessary to calculate the transformed functions~$|{\rm \chi}\rangle$.
Our derivation is restricted to the transformation of~$1S_{1/2}$ state of relativistic hydrogen-like
atom, its generalization to other states is straightforward.

\subsection{Discretization of continuum states}

Consider an arbitrary state~$|{\rm n\rangle}$ of~$\hH$.
The bound states~$|{\rm n}\rangle$ depend on the integer quantum number~$n$,
while continuum states depend on the absolute value of electron's
momentum~$p=\sqrt{(E/c)^2-(m_0c)^2}$, and
the sign of energy branch~$\epsilon=\pm 1$. Three other quantum numbers describing~$|{\rm n}\rangle$,
namely the total angular momentum~$j$, the orbital number~$l$ and~$m=j_z$,
are omitted in the present section. We assume that the bound states are normalized
to the Kronecker delta:~$\langle n_1|n_2\rangle = \delta_{n_1,n_2}$, while the continuum states are
normalized to the Dirac
delta~$\langle p_1^{\epsilon_i}|p_2^{\epsilon_j}\rangle = \delta(p_1-p_2)\delta_{\epsilon_i,\epsilon_j}$.

Let~$\psi_{1S}({\bm r})= \langle {\bm r}|\text{1S}\rangle$ denotes the~$1S_{1/2}$
state of the relativistic hydrogen-like atom with spin up.
Then writing~$|{\rm n}_0\rangle =|\text{1S}\rangle$ in Eq.~(\ref{ET_chi}) one obtains
\begin{equation} \label{DISC_0}
 \chi_{1S}({\bm r}) \equiv \langle {\bm r}|\hU|\text{1S}\rangle = \sum_n a_n\psi_{n}({\bm r}) +
 \sum_{\epsilon=\pm}\int_0^{\infty}\! a_p^{\epsilon}\psi_p^{\epsilon}({\bm r}),
\end{equation}
where~$a_n=\langle \psi_n|\chi_{1S}\rangle$ and~$a_p^{\epsilon}=\langle \psi_p^{\epsilon}|\chi_{1S}\rangle$.
The difficulty is caused by the different normalization of bound and
continuum states. To surmount this problem we follow Refs.~\cite{BransdenBook,Kocbach2001} by
replacing the continuum functions~$\psi_p^{\epsilon}({\bm r})$
in Eq.~(\ref{DISC_0}) by the so-called discretized functions~$\Psi_{p_i}^{\epsilon}({\bm r})$
\begin{eqnarray}
\chi_{1S}({\bm r}) &\simeq& \sum_n a_n\psi_{n}({\bm r}) + \sum_{\epsilon=\pm} \nonumber\\
&& \sum_{p_i}\left(a_{p_i}^{\epsilon} \sqrt{\Delta p} \right)
\label{DISC_1}
\left(\frac{1}{\sqrt{\Delta p}} \int_{p_i-\frac{\Delta p}{2}}^{p_i+\frac{\Delta p}{2}} \psi_p^{\epsilon}({\bm r}) dp \right) \ \ \\
\label{DISC_2}
&=& \sum_n a_n\psi_{n}({\bm r}) + \sum_{\epsilon=\pm} \sum_{p_i} A_{p_i}^{\epsilon} \Psi_{p_i}^{\epsilon}({\bm r}),
\end{eqnarray}
where~$A_{p_i}^{\epsilon}$ and~$\Psi_{p_i}^{\epsilon}({\bm r})$ are defined by the first and
second brackets in Eq.~(\ref{DISC_1}), respectively.
As shown in Refs.~\cite{BransdenBook,Bertulani1992,Kocbach2001}, for~$\Delta p \rightarrow 0$
the summation over~$p_i$ in Eq.~(\ref{DISC_2}) reduces to the integration over~$dp$,
as given in Eq.~(\ref{DISC_0}).
The essential features of functions~$\Psi_{p_i}^{\epsilon}({\bm r})$ are:
a) they are localized and integrable in the real space~\cite{Bertulani1992};
b) they are normalized to the Kronecker
delta:~$\langle \Psi_{p_i}^{\epsilon_i}|\Psi_{p_j}^{\epsilon_j}\rangle = \delta_{p_i,p_j}\delta_{\epsilon_i,\epsilon_j}$;
c) they are orthogonal to all functions~$\psi_{n}$ of bound states;
d)~$\Psi_{p_i}^{\epsilon_i}$ and~$\psi_{n}$ form a complete basis for four-component spinors.
Therefore, the bound states~$\psi_{n}({\bm r})$ and discretized functions~$\Psi_{p_i}^{\epsilon}({\bm r})$
can be treated similarly, i.e., all integrals including continuum functions~$\psi_p({\bm r})$ may be
replaced by sums over discretized functions~$\Psi_{p_i}^{\epsilon}({\bm r})$. The discretized functions in Eq.~(\ref{DISC_2})
are also called in literature eigendifferentials~\cite{GreinerIntro}.

\subsection{Probability amplitudes}

To find the probability amplitudes~$a_n$ and~$A_{p_i}^{\epsilon}$ in Eq.~(\ref{DISC_2})
we calculate the matrix elements of~$\hU$ between eigenstates~$1S_{1/2}$ and~$|n\rangle$
\begin{eqnarray}
\label{EC_an}
 a_n                &=& \langle n| \hU|\text{1S}\rangle = \langle n|\hZ|\text{1S}\rangle + \langle n|\hZ\hB\hl|\text{1S}\rangle, \\
 \label{EC_Api}
 A_{p_i}^{\epsilon} &=& \langle\Psi_{p_i}^{\epsilon}|\hU|\text{1S}\rangle = \langle\Psi_{p_i}^{\epsilon}|\hZ|\text{1S}\rangle +
                        \langle\Psi_{p_i}^{\epsilon}|\hZ\hB\hl|\text{1S}\rangle,
\end{eqnarray}
with the normalization condition
\begin{equation} \label{EC_sumAa}
 \sum_n|a_n|^2 + \sum_{\epsilon = \pm} \sum_{p_i} \left|A_{p_i}^{\epsilon}\right|^2 = 1.
\end{equation}

The operator~$\hZ$ has vanishing matrix elements between the eigenstates of~$\hH$
having positive and negative energies, see Appendix~\ref{Appendix_Z}.
For an arbitrary eigenstate~$|{\rm n}\rangle$
there is~$\hl|\text{1S}\rangle= |\text{1S}\rangle$, see Eq.~(\ref{ET_Theta}).
The use of Eq.~(\ref{ET_Theta}) leads to the exact
treatment of the sign operator~$\hl$ in the calculation of the matrix elements of~$\hU$.
Inserting the unity operator
\begin{equation} \label{EC_one}
 \hat{1} = \sum_{n}|n\rangle \langle n| + \sum_{\epsilon=\pm 1} \sum_{p_i} |\Psi_{p_i}^{\epsilon}\rangle \langle \Psi_{p_i}^{\epsilon}|
\end{equation}
into expressions for~$a_n$ and~$A_{p_i}^{\epsilon}$ in Eqs.~(\ref{EC_an})--(\ref{EC_Api}), we obtain
\begin{eqnarray}
\label{EC_ans}
a_n = &\langle n|\hZ|\text{1S}\rangle +& \sum_{n'}\langle n|\hZ|n'\rangle \langle n'|\hB|\text{1S}\rangle + \nonumber \\
      &&+ \sum_{p_j}\langle n|\hZ|\Psi_{p_j}^+\rangle \langle \Psi_{p_j}^+|\hB|\text{1S}\rangle, \\
\label{EC_App}
 A_{p_i}^+ =& \langle \Psi_{p_i}^+|\hZ|\text{1S}\rangle &+
    \sum_{n'}\langle \Psi_{p_i}^+|\hZ|n'\rangle \langle n'|\hB|\text{1S}\rangle + \nonumber \\
    &&+ \sum_{p_j} \langle \Psi_{p_i}^+|\hZ|\Psi_{p_j}^+\rangle \langle \Psi_{p_j}^+|\hB|\text{1S}\rangle, \\
 \label{EC_Amm}
 A_{p_i}^-&=& +\sum_{p_j} \langle \Psi_{p_i}^-|\hZ|\Psi_{p_j}^-\rangle \langle \Psi_{p_j}^-|\hB|\text{1S}\rangle.
\end{eqnarray}
Thus, to find the probability amplitudes~$a_n$ and~$A_{p_i}^{\epsilon}$ one needs to calculate seven
types of matrix elements listed below
\begin{eqnarray}
\label{EC_lista}
\langle n'|\hB|n\rangle, \hspace*{1ex} \langle \Psi_{p_i}^+|\hB|n\rangle, \hspace*{1ex} \langle \Psi_{p_i}^-|\hB|n\rangle, \nonumber \\
\langle n'|\hZ|n\rangle, \hspace*{1ex} \langle \Psi_{p_i}^+|\hZ|n\rangle, \nonumber \\
\langle \Psi_{p_i}^+|\hZ|\Psi_{p_j}^+\rangle, \hspace*{1ex} \langle \Psi_{p_i}^-|\hZ|\Psi_{p_j}^-\rangle.
\end{eqnarray}

The matrix elements of~$\hB$ between two eigenstates of~$\hH$
are obtained in a standard way by computing four integrals between the components of these states, see below.
A calculation of matrix elements of~$\hZ$ is more complicated since this operator can not be represented
in a closed form, see Eq.~(\ref{ET_hZ}). To find~$\hZ$ in Eq.~(\ref{ET_hZ}) we introduce an auxiliary operator
\begin{equation} \label{HS_Sl}
 \hS = \hl\hB +\hB\hl = \sum_{\rm s} \xi_{\rm s}|{\rm s} \rangle \langle {\rm s}|,
\end{equation}
where~$\xi_{\rm s}$ and~$|{\rm s} \rangle$ are the eigenvalues and eigenvectors of~$\hS$, respectively.
Then, assuming that the square root exists, one has
\begin{equation} \label{HS_Zl}
 \hZ = \sum_{\rm s} \frac{1}{\sqrt{2+\xi_{\rm s}}}|{\rm s} \rangle \langle {\rm s}|.
\end{equation}

\begin{figure}
\includegraphics[width=8cm,height=8.0cm]{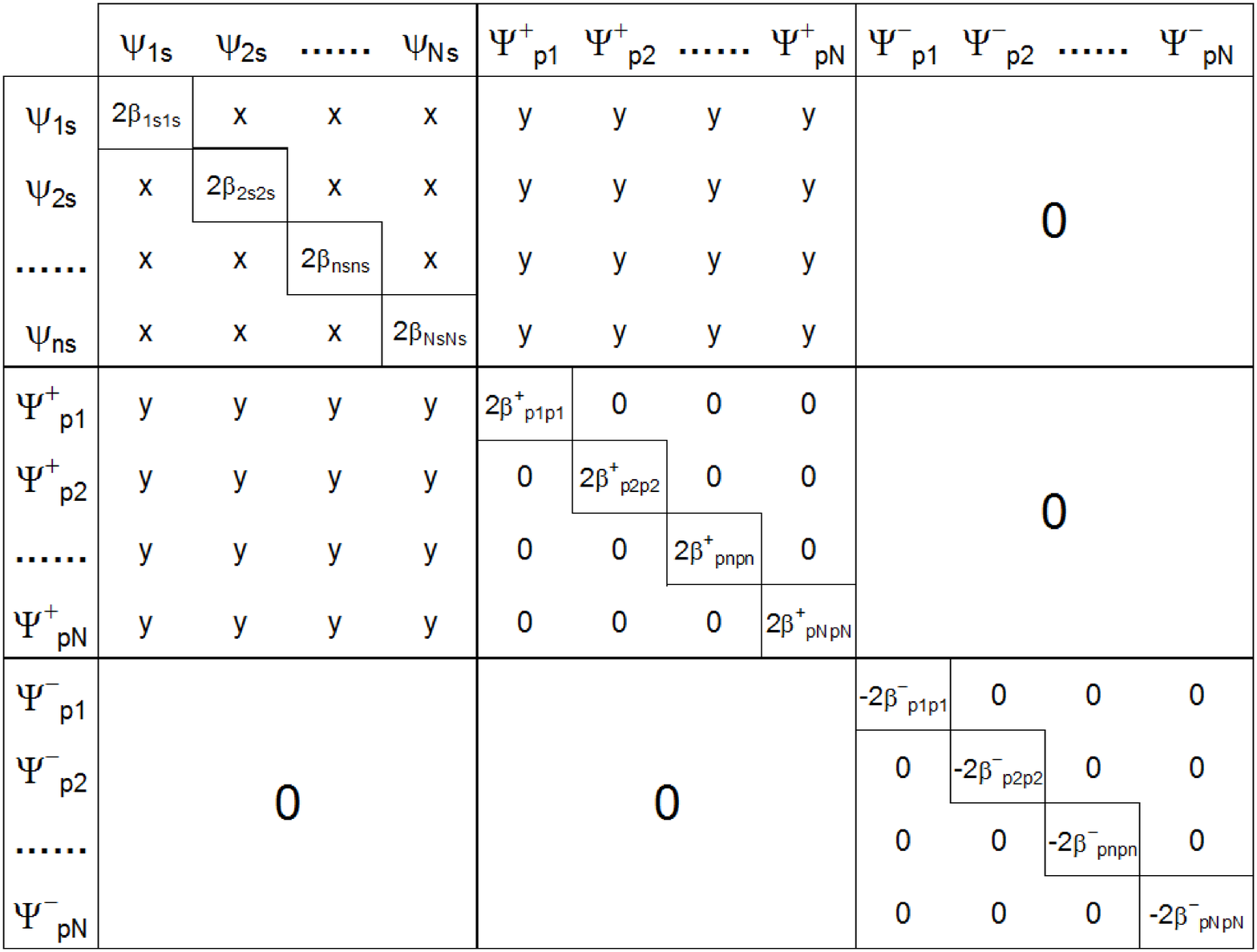}
\caption{Schematic matrix~$\hS=\hB\hl+\hl\hB$ in the basis of
         states~$\psi_{nS}$ and discretized functions~$\Psi_{p_i}^{\pm}$ in Eq.~(\ref{DISC_2}).
         Diagonal matrix elements, which dominate over nondiagonal ones, are shown explicitly.
         Nondiagonal elements between bound states are indicated by~$x$,
         while nondiagonal elements between bound and continuum states are indicated
         by~$y$. Note the block-diagonal form of matrix~$\hS$ and vanishing matrix
         elements between states of positive and negative energies.} \label{Fig1}
\end{figure}

Equations~(\ref{HS_Sl}) and~(\ref{HS_Zl}) give us a practical way for calculating~$\hZ$
in four steps. We first select the basis
consisting of three sets of eigenstates of~$\hH$: bound states~$\{\psi_n\}$,
discretized states having positive energies~$\{\Psi_{p_i}^+\}$,
and discretized states having negative energies~$\{\Psi_{p_i}^-\}$. Then
we calculate the matrix elements of~$\hS$ in Eq.~(\ref{HS_Sl}) between these states.
The resulting matrix has a block-diagonal form plotted schematically in Fig.~1.
The three sets of states composing the basis of matrix~$\hS_{{\rm n}_1,{\rm n}_2}$
are infinite and the matrix in Fig.~1 is of infinite order. In our calculations we truncate the
infinite sets to finite ones, see Table~I.
In the third step one calculates all eigenvalues and eigenstates of~$\hS$, as
given in Eq.~(\ref{HS_Sl}), and in the last step one constructs the operator~$\hZ$ in Eq.~(\ref{HS_Zl}).
After this process~$\hZ$ is approximated by a finite matrix of the
size~$=1576\times 1576$, see Table~I.

\begin{table}
\label{Table_Par}
\caption{Model parameters used in calculations}
\begin{tabular}{|l|c|}
\hline
 Quantity & Value \\
 \hline
 Number of bound states                &  40  \\
 Number of continuum states for~$E >0$ & 512  \\
 Number of continuum states for~$E <0$ & 1024 \\
 \hline
 \multicolumn{2}{|c|}{Summation over discretized states in Eqs.~(\ref{DISC_1}) and~(\ref{DISC_2})} \\
 \hline
 $\Delta p$                      &   0.1~$\hbar(r_B/Z)^{-1}$ \\
 lower limit of~$p_i$            &   0.1~$\hbar(r_B/Z)^{-1}$ \\
 upper limit of~$p_i$ for~$E >0$ &  51.2~$\hbar(r_B/Z)^{-1}$ \\
 upper limit of~$p_i$ for~$E <0$ & 102.4~$\hbar(r_B/Z)^{-1}$ \\
 \hline
\end{tabular}
\end{table}

\subsection{Foldy-Wouthuysen-like approximation for~\mbox{$\hU$}}

Following Eriksen, one can formally expand the square root in Eq.~(\ref{ET_hZ})
treating~$\hat{q}= \hB\hl - 1$ as an expansion parameter.
In the second order in~$\hat{q}$, one obtains~\cite{Eriksen1958}
\begin{equation} \label{FW_U1}
 \hU \simeq 1 + \frac{1}{2}\hat{q} - \frac{1}{8}\left( \hat{q} + \hat{q}^*\right) + \ldots.
\end{equation}
Eriksen showed that this expansion is equivalent to the FW transformation
in the second order of~$1/(m_0c^2)$.
Functions transformed with the use of Eriksen and FW-transformations have similar upper components.
For this reason we shall refer to the transformation
defined in Eq.~(\ref{FW_U1}) as the FW-like expansion of Eriksen operator.
After simple algebra one obtains
\begin{equation} \label{FW_U2}
\hU^{FW} = \frac{3}{4} + \frac{1}{2} \hB\hl - \frac{1}{8}\left( \hB\hl + \hl\hB \right).
\end{equation}
The operator~$\hU^{FW}$ in Eq.~(\ref{FW_U2}) is a linear function of~$\hS=\hB\hl+\hl\hB$,
and it is not necessary to calculate its inverted square root~$\hZ$,
see Eqs.~(\ref{ET_hZ}) and~(\ref{HS_Zl}). This simplifies calculations of
transformed functions~$\hU^{FW}|\psi\rangle$.

Let~$|{\rm n}_1\rangle$,~$|{\rm n}_2\rangle$ be two eigenstates of~$\hH$,
whose signs are~$\epsilon_1$ and~$\epsilon_2$, respectively. For~$\epsilon_1=\epsilon_2$
we have
\begin{equation} \label{FW_12a}
 \langle {\rm n}_1|\hU^{FW}|{\rm n}_2\rangle = \frac{3}{4}\delta_{{\rm n}_1,{\rm n}_2} + \frac{1}{4}\epsilon_2 \hB_{{\rm n}_1,{\rm n}_2},
\end{equation}
while for~$\epsilon_1=-\epsilon_2$ there is
 \begin{equation} \label{FW_12b}
\langle {\rm n}_1|\hU^{FW}|{\rm n}_2\rangle = \frac{1}{2} \epsilon_2 \hB_{{\rm n}_1,{\rm n}_2}.
\end{equation}
Comparing Eqs.~(\ref{EC_ans})--(\ref{EC_Amm}) and~(\ref{FW_12a})--(\ref{FW_12b})
we see that in both cases one has to calculate the same matrix elements of~$\hB$, as
listed in Eq.~(\ref{EC_lista}).

\section{Eigenstates of~\mbox{$\hH$} and selection rules for~\mbox{$\hB_{\lowercase{\rm n}_1,\lowercase{\rm n}_2}$}} \label{Sec_SE}

In this section we find selection rules for the matrix elements of~$\hB$ between eigenstates of~$\hH$.
Next we introduce the radial wave functions of bound and continuum states of~$\hH$.
We also qualitatively estimate magnitudes of radial integrals used in the matrix elements of~$\hB$.

\subsection{Selection rules for~\mbox{$\hB_{{\rm n}_1,{\rm n}_2}$}}

Because of spherical symmetry of the problem,
the eigenstates~$\langle {\bm r}|{\rm n}\rangle$
are products of radial functions~$g(r)$ and~$f(r)$,
and function~$\Omega(\theta,\varphi)$ depending on angular variables.
The latter is characterized by three quantum numbers: orbital angular momentum~$l$,
total angular momentum~$j$ and~$m=j_z$. The auxiliary quantum number~$\kappa$ is defined as
\begin{equation} \label{SR_kappa}
 \kappa = \left\{\begin{array}{cc}
            -j - \frac{1}{2} & \text{for}\ \ j=l+\frac{1}{2}, \\
             j + \frac{1}{2} & \text{for}\ \ j=l-\frac{1}{2}. \end{array} \right.
\end{equation}
It is either positive or negative integer, but not zero.
The eigenstates of~$\hH$ are~\cite{RoseBook,GreinerBook}
\begin{equation} \label{SR_psi}
\psi_{\rm n}(r,\theta,\varphi) =
 \left( \begin{array}{c}
   g(r)\ \Omega_{\kappa, m}(\theta,\varphi) \\
  if(r)\ \Omega_{\kappa, -m}(\theta,\varphi)
   \end{array} \right),
\end{equation}
where~$\Omega_{\kappa, m}(\theta,\varphi)$ are two-component spinors:
\begin{equation} \label{SR_Omega}
\Omega_{\kappa, m}(\theta,\varphi) =
 \left( \begin{array}{c}
 \ \ \ \ \ \ \sqrt{\frac{\kappa+\frac{1}{2}-m}{2\kappa+1}}\ Y_{\kappa, m-1/2}(\theta,\varphi) \\
 -\frac{\kappa}{|\kappa|}\sqrt{\frac{\kappa+\frac{1}{2}+m}{2\kappa+1}}\ Y_{\kappa, m+1/2}(\theta,\varphi)
 \end{array} \right),
\end{equation}
and~$Y_{a,b}(\theta,\varphi)$ are the spherical harmonics. For~$a>0$ the latter are defined as
\begin{equation} \label{SR_Yab}
Y_{a,b}(\theta,\varphi) = (-1)^b \sqrt{\frac{2a+1}{4\pi}\frac{(a-b)!}{(a+b)!}} P_a^b(\cos(\theta))e^{ib\varphi},
\end{equation}
where~$P_a^b(\cos(\theta))$ are the associated Legendre polynomials in the usual notation~\cite{GradshteinBook}. For~$a<0$
there is~$Y_{a,b}(\theta,\varphi) = Y_{-a,-b}(\theta,\varphi)$, and for~$|a| < |b|$ there is~$Y_{a,b}(\theta,\varphi)\equiv 0$.
The functions~$\Omega_{\kappa, m}(\theta,\varphi)$ fulfill the orthogonality relation~\cite{GreinerBook}
\begin{eqnarray} \label{SR_ort}
 \int_0^{\pi} \int_0^{2\pi} \Omega_{\kappa_1, m_1}(\theta,\varphi)\Omega_{\kappa_2, m_2} (\theta,\varphi)\sin(\theta)d\theta d\varphi\ \nonumber \\
 =\delta_{\kappa_1\kappa_2}\delta_{m_1m_2}.
\end{eqnarray}

The matrix element of~$\hB$ between two eigenstates~$|{\rm n}_1\rangle$ and~$|{\rm n}_2\rangle$ is
\begin{eqnarray} \label{SR_kkmm}
\hB_{{\rm n}_1,{\rm n}_2} &=& \int d^3{\bm r}
                            \left( \begin{array}{c}g_1\Omega_{\kappa_1,m_1} \\ if_1 \Omega_{\kappa_1,-m_1} \end{array} \right)^{\dagger}
                            \left( \begin{array}{cc} \hat{1} & 0 \\ 0 & -\hat{1} \end{array} \right)
                            \left( \begin{array}{c}g_2\Omega_{\kappa_2,m_2} \\ if_2 \Omega_{\kappa_2,-m_2} \end{array} \right) \nonumber \\
                         &=& \delta_{\kappa_1\kappa_2}\delta_{m_1m_2}\int_0^{\infty}\left( g_1 g_2 - f_1 f_2 \right) r^2 dr.
\end{eqnarray}
Equation~(\ref{SR_kkmm}) gives selection rules for the matrix elements of~$\hB$ between eigenstates of~$\hH$.
The matrix element~$\hB_{{\rm n}_1,{\rm n}_2}$ is nonzero only for states having the same
quantum numbers~$\kappa$ and~$m$. Returning to Eqs.~(\ref{EC_ans})--(\ref{EC_Amm})
we see that all intermediate states~$|n'\rangle$ and~$|\Psi_{p_j}^{\pm}\rangle$ must
be described by the same quantum numbers~$\kappa=-1$ and~$m=1/2$
as the initial state~$|\text{1S}\rangle$.
This holds both for the bound and continuum eigenstates of~$\hH$.
Below we analyze three groups of states characterized by quantum
numbers listed in Table~II. An extension of the results to states
described by other quantum numbers is straightforward.

\begin{table}
\label{Table_Ladd}
\caption{Quantum numbers characterizing eigenstates of relativistic hydrogen-like
         atom used for calculations in Figs.~2 to~6.}
\begin{tabular}{|c|c|c|c|c|}
\hline
Eigenstate &~$\kappa$ &~$m$ &~$n_r$ & Examples of intermediate states\\
 \hline
 $1S_{1/2}$ & -1 & 1/2 & 0 & $1S_{1/2}$,~$2S_{1/2}$,~$3S_{1/2}$,\ldots\\
 $2P_{1/2}$ &  1 & 1/2 & 1 & $2P_{1/2}$,~$3P_{1/2}$,~$4P_{1/2}$,\ldots\\
 $2P_{3/2}$ & -2 & 3/2 & 1 & $2P_{3/2}$,~$3P_{3/2}$,~$4P_{3/2}$,\ldots\\
 \hline
\end{tabular}
\end{table}

\subsection{Radial wave functions}

To complete determination of the matrix elements~$\hB_{{\rm n}_1,{\rm n}_2}$ in Eq.~(\ref{SR_kkmm}) we
need to calculate the integrals over the radial functions~$g(r)$ and~$f(r)$. These functions are known
explicitly for both bound and continuum states. The integrals in Eq.~(\ref{SR_kkmm}) can be computed
either numerically or analytically, see below. In this work we use analytical expressions for the integrals,
which is more accurate for large quantum numbers~$n$ or large values of electron momentum.

For bound eigenstates the functions~$g(r)$ and~$f(r)$ are~\cite{AkhiezerBook,GreinerBook}
\begin{eqnarray}
 \label{RF_rgd}
 rg(r) &=& - {\cal A}\ \sqrt{1+W_{n_r}}\ {\cal B}\ \left\{n_r {\cal F}_1 - (N-\kappa){\cal F}_0 \right \}, \\
 \label{RF_rfd}
 rf(r) &=& - {\cal A}\ \sqrt{1-W_{n_r}}\ {\cal B}\ \left\{n_r {\cal F}_1 + (N-\kappa){\cal F}_0 \right \},
\end{eqnarray}
where
\begin{eqnarray}
 \label{RF_A}
 {\cal A} &=& \frac{\sqrt{\Gamma(2\gamma+1+n_r)}}{\Gamma(2\gamma+1)\sqrt{4N(N-\kappa)n_r!}}\left(\frac{2Z}{Nr_B} \right)^{1/2}, \\
 \label{RF_2}
 {\cal B} &=& \exp\left(-\frac{Zr}{Nr_B} \right) \left(\frac{2Zr}{Nr_B} \right)^{\gamma}, \\
 \label{RF_C}
 {\cal F}_{\nu} &=& {_1}F_1\left(-n_r+\nu,2\gamma+1, \frac{2Zr}{Nr_B} \right),
\end{eqnarray}
where~$\nu=0,1$. The discrete energy is
\begin{equation} \label{RF_En}
 E_{n_r}= m_0c^2 \left\{ 1 + \left(\frac{\alpha Z}{n_r+\gamma}\right)^2 \right\}^{-1/2},
\end{equation}
in which~$\gamma=\sqrt{\kappa^2-(\alpha Z)^2}$,~$\alpha=e^2/(4\pi\epsilon_0 \hbar c) \simeq 1/137$ is
the fine-structure constant,~$W_{n_r}=E_{n_r}/m_0c^2< 1$,
~$N=\sqrt{n^2-2n_r(|\kappa|-\gamma)}$,~$n=n_r+|\kappa|$,~$r_B\simeq 0.51 \AA$ is the Bohr radius,
and the function~$_1F_1(a,c,z)$ is the confluent hypergeometric function in the standard
notation~\cite{GradshteinBook}. Since the first argument of the confluent hypergeometric function
in Eq.~(\ref{RF_C}) is negative integer, this function reduce to polynomial
of the order~$n_r-\nu$ of~$z=2Zr/(Nr_B)$. Expanding~$E_{n_r}$ in Eq.~(\ref{RF_En})
in the vicinity of~$m_0c^2$ one obtains
\begin{equation} \label{RF_W}
 W_{n_r} = \frac{E_{n_r}}{m_0c^2} \simeq 1 -\frac{(\alpha Z)^2}{2(n_r+|\kappa|)^2} + \ldots.
\end{equation}
Thus the functions~$g(r)$ in Eq.~(\ref{RF_rgd}) are on the order of unity,
while the functions~$f(r)$ in Eq.~(\ref{RF_rgd}) are on the order of~$\alpha Z < 1$.

Let~$|1\rangle$ and~$|2\rangle$ be two eigenstates of~$\hH$ with the same quantum
numbers~$\kappa$ and~$m$. Then
\begin{eqnarray}
 \label{RF_n1n2}
 \langle 1|    2\rangle =& \int_0^{\infty} [g_1(r)g_2(r) + f_1(r)f_2(r)]r^2 dr =& \delta_{1,2}, \\
 \label{RF_bn1n2}
 \langle 1|\hB|2\rangle =& \int_0^{\infty} [g_1(r)g_2(r) - f_1(r)f_2(r)] dr =& \hB_{12}.
\end{eqnarray}
Subtracting the above equations one finds
\begin{equation} \label{RF_b12}
 \hB_{12} = \delta_{1,2} - 2 \int_0^{\infty} f_1(r)f_2(r) r^2 dr.
\end{equation}
Since in Eq.~(\ref{RF_b12}) the functions~$f_1(r)$ and~$f_2(r)$ are on the order of~$\alpha Z$,
the diagonal matrix elements of~$\hB$ are on the order of unity,
while the nondiagonal ones are on the order of~$(\alpha Z)^2 \ll 1$.

Continuum radial functions~$g(r)$ and~$f(r)$ are~\cite{GreinerBook,Rose1937,RoseBook}
\begin{eqnarray}
 \label{RF_rgc}
 rg(r) &=& +        \sqrt{(|W_p|+\epsilon)} \left( {\cal D}{\cal F} + {\cal D}^*{\cal F}^* \right), \\
 \label{RF_rfc}
 rf(r) &=& i\epsilon\sqrt{(|W_p|-\epsilon)} \left( {\cal D}{\cal F} - {\cal D}^*{\cal F}^* \right).
\end{eqnarray}
The electron energy is
\begin{equation}\label{RF_Ep}
 E_p= \epsilon\sqrt{(m_0c^2)^2 + (cp)^2},
\end{equation}
Here~$\epsilon=\pm 1$ is the energy sign,~$W_p=E_p/(m_0c^2)$ and~$|W_p| > 1$. Then
\begin{eqnarray}
\label{RF_D}
{\cal D} &=& \frac{e^{\pi y/2}|\Gamma(\gamma+iy)|}{2(\pi|W_p|)^{1/2}\Gamma(2\gamma+1)}
\left[ e^{i\eta}(\gamma+iy)\right], \\
 \label{RF_F}
{\cal F} &=& (2kr)^{\gamma}e^{-ikr} {_1}F_1(\gamma+1+iy,2\gamma+1,2ikr).
\end{eqnarray}
The momentum of relativistic electron is~$\hbar k = (\hbar/\lambda_c)\sqrt{W_p^2-1} >0$,
where~$\lambda_c=\hbar/(m_0c)$ is the Compton wavelength. Finally~$y=\alpha Z W_p/\sqrt{W_p^2-1}$ and
\begin{equation} \label{RF_eieta}
 e^{i\eta} = \left(-\frac{\kappa - iy/W_p}{\gamma+iy}\right)^{1/2}.
\end{equation}
Function~$\psi_p^{\epsilon}(r) = \left(\begin{array}{c} g(r) \\ f(r) \end{array} \right)$ is
normalized to~$\delta(p_1-p_2)\delta_{\epsilon_1\epsilon_2}$.
The asymptotic forms of~$g(r)$ and~$f(r)$ are
\begin{eqnarray}
 \label{RF_rgAS}
 rg(r) &\simeq& +        \sqrt{\frac{|W_p|+\epsilon}{\pi |W_p|}} \cos(kr+\delta) \\
 \label{RF_frAS}
 rf(r) &\simeq& -\epsilon\sqrt{\frac{|W_p|-\epsilon}{\pi |W_p|}} \sin(kr+\delta),
\end{eqnarray}
where
\begin{equation} \label{RF_delta}
 \delta= y\ln(2kr) -\arg[\Gamma(\gamma+iy)] - \frac{\pi\gamma}{2} + \eta.
\end{equation}
Thus, for large arguments,~$g(r)$ and~$f(r)$ reduce to trigonometric functions
with slowly varying phases. For small electron momenta there is:~$|W_p| \simeq 1 + p^2/(2m_0^2c^2)$.
Then, for~$\epsilon=+1$, the function~$g(r)$ in Eq.~(\ref{RF_rgc}) is on the order of unity,
while~$f(r)$ in Eq.~(\ref{RF_rfc}) is on the order of~$p/(m_0c) \ll 1$.
For~$\epsilon=-1$ the magnitudes of~$g(r)$ and~$f(r)$ reverse:~$f(r)$ is on the
order of unity while~$g(r)$ is on the order of~$p/(m_0c) \ll 1$.
For large electron momenta:~$p \ge m_0c$, the magnitudes of~$g(r)$ and~$f(r)$ are similar.

Let us qualitatively estimate magnitudes of the matrix elements~$\hB_{nS,p\epsilon}$ between the
bound state~$|nS\rangle$ and the continuum state~$|p\ \epsilon\rangle$ of small momentum,
described by the same values of~$\kappa$ and~$m$. Combining the above estimations with those
for the functions of bound states [see Eq.~(\ref{RF_W})] we have
\begin{equation} \label{RF_ld}
\hB_{nS,p\epsilon} \propto
 \left\{ \begin{array}{cc} \alpha Z \times p/(m_0c) & \text{for}\ \epsilon=+1, \\
         \alpha Z & \text{for}\ \epsilon=-1. \end{array}
 \right.
\end{equation}
Since~$p/(m_0c) \ll 1$, the continuum states having negative energies are expected
to give much larger contributions to the transformed functions than the continuum
states having positive energies. This conclusion is confirmed numerically below.

The matrix elements of~$\hB$ between radial wave functions can be calculated analytically
or numerically. There are several works related to this subject, see e.g.
Refs.~\cite{Goldman1982,Suslov2009,Reynolds1964,Gargaro1970}.
For the analytical results for the diagonal matrix elements,
we refer the reader to Appendix~\ref{Appendix_Beta}.

\section{Results} \label{Sec_Res}

\begin{figure}
\includegraphics[width=8.0cm,height=8.0cm]{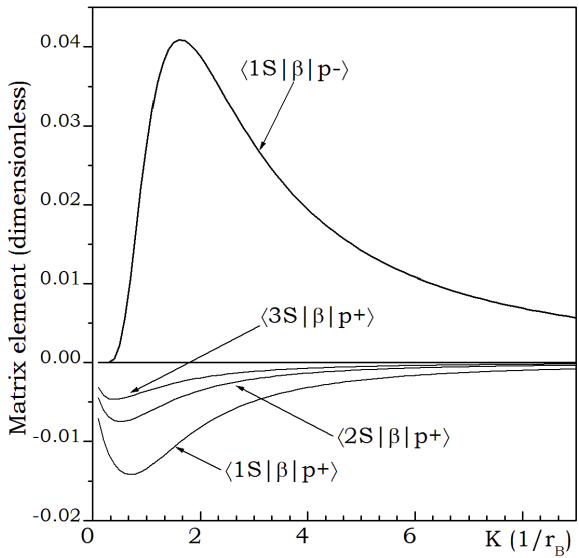}
\caption{Matrix elements~$\hB_{nS,p\pm}$ versus effective wave vector~$K=(p/\hbar)(\alpha Z)$
         for several values of~$n$. Results for~$n=1$ are obtained analytically
         from Eq.~(\ref{AppMEB_Q2}).} \label{Fig2}
\end{figure}

\begin{figure}
\includegraphics[width=8.0cm,height=8.0cm]{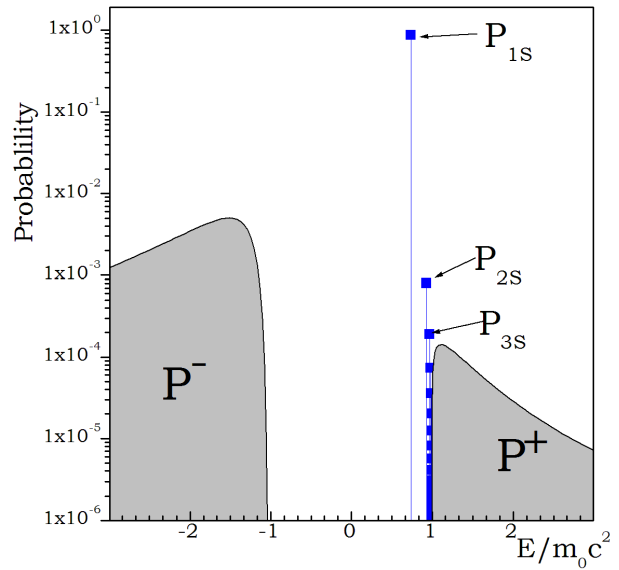}
\caption{Probability densities~$|a_n|^2$ and~$|A_{p_i}^{\pm}|^2$ calculated from Eqs.~(\ref{EC_ans})--(\ref{EC_Amm})
         for~$Z=92$ versus electron energy.
         The probability densities~$|a_n|^2$ for three lowest~$n$ are shown explicitly.
         The shaded areas indicate integrated probability densities~$P^{\pm}$ for continuum states
         of positive and negative energies, see Fig.~4 for~$Z=92$.} \label{Fig3}
\end{figure}

Our calculations were performed numerically
for spin-up states~$1S_{1/2}$,~$2P_{1/2}$ and~$2P_{3/2}$ for~$Z \in \{1 \ldots 92\}$,
but for brevity we quote results for five representative values of~$Z$.
In Table~I we list numerical and model parameters used in the calculations.

We begin with the analysis of matrix elements of~$\hB$ between bound states.
In Table~III we show values of~$\hB_{1S,ns}$ for several~$Z$ and~$n$.
For all~$Z$ the diagonal elements~$\hB_{1S,1S}$ dominate over the nondiagonal ones,
which agrees with qualitative estimations in Eq.~(\ref{RF_b12}).
The nondiagonal elements of~$\hB$ gradually decay with~$n$, and the analysis indicates that they vanish as~$1/n^{3/2}$.
For small~$Z$ the diagonal elements are nearly equal to unity, while the nondiagonal ones
are negligible. With increasing~$Z$ the diagonal elements gradually decrease and
other bound states begin to be relevant.

\begin{table}
\label{Table_Beta}
\caption{Matrix elements of~$\hB$ between bound states of relativistic hydrogen-like atom
         for several values of the nuclear charge~$Z$.}
\begin{tabular}{|c|c|c|c|c|c|}
  \hline
  Z & Element & $\hB_{1S,1S}$ & $\hB_{1S,3S}$ & $\hB_{1S,5S}$ & $\hB_{1S,10S}$\\
  \hline
  1  & H  & 0.99 & -3.7$\times$ 10$^{-6}$ & -2.0$\times$ 10$^{-6}$ & -7.9$\times$ 10$^{-7}$ \\
  24 & Cr & 0.98 & -2.1$\times$ 10$^{-3}$ & -1.1$\times$ 10$^{-3}$ & -4.6$\times$ 10$^{-4}$ \\
  47 & Ag & 0.94 & -8.2$\times$ 10$^{-3}$ & -4.4$\times$ 10$^{-3}$ & -1.7$\times$ 10$^{-3}$ \\
  74 & W  & 0.84 & -2.0$\times$ 10$^{-2}$ & -1.1$\times$ 10$^{-2}$ & -4.2$\times$ 10$^{-3}$ \\
  92 & U  & 0.74 & -3.2$\times$ 10$^{-2}$ & -1.6$\times$ 10$^{-2}$ & -6.4$\times$ 10$^{-3}$ \\
  \hline
\end{tabular}
\end{table}

Next we calculate the matrix elements (MEs) of~$\hB_{nS,p\epsilon}$ between bound states~$|nS\rangle$
and continuum ones~$|p\ \epsilon\rangle$ as functions of the effective wave vector~$K=(p/\hbar)(\alpha Z)$.
In Fig.~2 we plot MEs~$\hB_{1S,p-}$ and~$\hB_{nS,p+}$ for~$Z=92$
and several values of~$n$. Elements~$\hB_{1S,p\pm}$ are calculated analytically
from Eq.~(\ref{AppMEB_Q2}).

Each curve in Fig.~2 has an asymmetric bell-like shape, it vanishes at~$K=0$,
has a maximum or minimum in the vicinity of~$K\simeq \alpha Z/(r_Bn)$, and decrease to zero for large~$K$.
The magnitudes of~$\hB_{1S,p-}$ are much larger than those of~$\hB_{nS,p+}$,
which underlines the greater contribution of negative energies states to~$\hU|1S\rangle$.
Finally, the magnitudes of~$\hB_{nS,p+}$
decreases with~$n$, so that only the first few states
give significant contributions to~$\hU|1S\rangle$.

Having calculated MEs~$\hB_{nS,mS}$ and~$\hB_{nS,p\pm}$ we compute the
probability amplitudes~$a_n$ and~$A_{p_i}^{\epsilon}$ and the resulting probability
densities~$|a_n|^2$ and~$|A_{p_i}^{\epsilon}|^2$, as given in Eqs.~(\ref{EC_an})--(\ref{EC_Api}).
The results for~$Z=92$ are shown in Fig.~3.
The state~$1S_{1/2}$ gives the largest contribution to~$\chi_{1S}$
and dominates over contributions of all other states.
But, surprisingly, the next significant contributions to~$\chi_{1S}$
originate from the continuum states of negative energies around~$E_p^- \simeq -1.5 m_0 c^2$.
The contributions for other states are negligible.
For the relativistic hydrogen atom~($Z=1$)
the corresponding probability densities are similar to those in Fig.~3,
but with a much larger contribution of the~$1S_{1/2}$ state, see below.

\begin{figure}
\includegraphics[width=8.0cm,height=8.0cm]{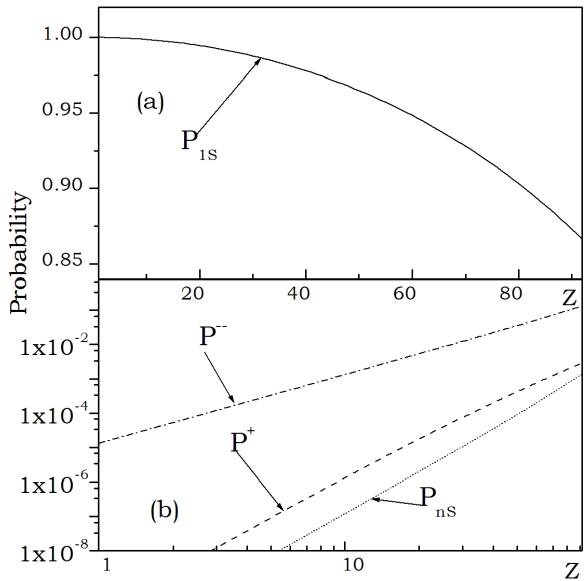}
\caption{Contribution of various eigenstates in transformed function~$\chi_{1S}$
         versus nuclear charge~$Z$.
         Solid line: probability density~$|a_1|^2$ for~$1S_{1/2}$ state.
         Dot-dashed line: integrated density~$P^-$ for continuum states of negative energies.
         Dashed line: integrated density~$P^+$ for continuum states of positive energies.
         Dotted line: integrated densities~$P_{nS}$ for bound states of~$\hH$
                      except~$1S_{1/2}$ state.} \label{Fig4}
\end{figure}

Let~$P_{1S}=|a_{1S}|^2$ denote the probability density for~$1S_{1/2}$ state of~$\hH$.
Let~$P^{\pm}=\sum_i|A_{p_i}^{\pm}|^2$ be the integrated probability density for
continuum states of~$\hH$, and~$P_{nS}=\sum_{n>1}|a_n|^2$
be the integrated probability density for bound states of~$\hH$ except~$1S_{1/2}$ state.
We plot these four quantities in Fig.~4 as functions of the nuclear charge~$Z \in [1\ldots 92]$.

For small~$Z$ the probability density~$P_{1S}$ is close to unity which means that, for light
atoms,~$\chi_{1S}$ is almost entirely composed of the state~$1S_{1/2}$.
For larger~$Z$ this probability density gradually decreases, but for~$Z=92$,
it is still~$P_{1S}\simeq 86\%$. Thus, even for very heavy atoms the transformed function
is composed mostly of the~$1S_{1/2}$ state.
For small~$Z$ all three integrated probability densities~$P^{\pm}$ and~$P_{nS}$ are negligible,
but for larger~$Z$ the~$P^-$ becomes about~$14\%$.
However, neither~$P^+$ or~$P_{nS}$ exceed~$0.1\%$ in the whole range of~$Z$,
so that their contributions to~$\chi_{1S}$ may be neglected.

\begin{figure}
\includegraphics[width=8.0cm,height=8.0cm]{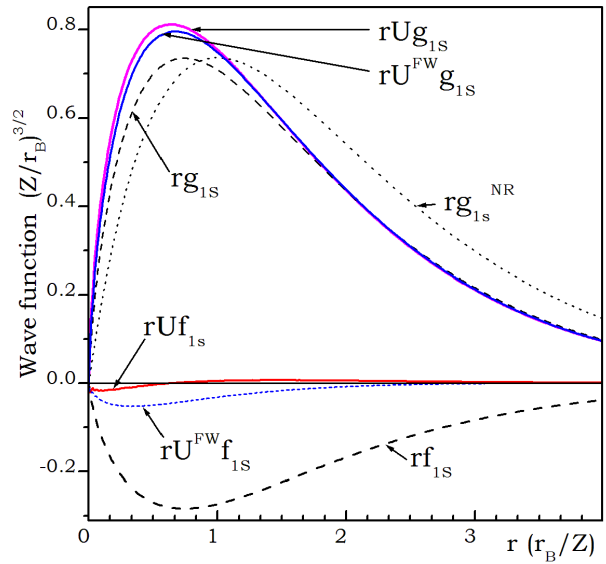}
\caption{Electron wave functions prior to and after Eriksen and FW-like transformations for~$1S_{1/2}$
         state of relativistic hydrogen-like atom with nuclear charge~$Z=92$.
         Negative values: functions~$rf_{1S}$,~$r\hU f_{1S}$ and~$r\hU^{FW} f_{1S}$.
         According to the Eriksen theory the function~$r\hU f_{1S}$ should be identically zero.
         The FW-like transformed function~$r\hU^{FW} f_{1S}$ is also close to zero,
         but its magnitude is larger than the magnitude of~$r\hU f_{1S}$. Positive values:
         functions~$rg_{1S}$,~$r\hU g_{1S}$,~$r\hU^{FW} g_{1S}$ and~$rg_{1S}^{NR}$.
         Note a small difference between
         exact~$r\hU g_{1S}$ and approximate~$r\hU^{FW} g_{1S}$ functions.} \label{Fig5}
\end{figure}

Having determined the expansion probability amplitudes~$a_n$ and~$A_{p_i}^{\pm}$
we calculate the transformed function~$\hU \psi_{1S}$ in Eq.~(\ref{ET_Un}).
In Fig.~5 we plot the results for~$Z=92$.
Functions~$rg_{1S}$ and~$rf_{1S}$ of the~$1S_{1/2}$ state
are marked by the dashed lines, while~$r\hU g_{1S}$ and~$r\hU f_{1S}$ of Eq.~(\ref{ET_hU})
are indicated by the solid lines. The dash-dotted lines are functions transformed with
use of~$\hU^{FW}$ given in Eq.~(\ref{FW_U2}). Finally, the dotted line
shows~$1s$ function of the nonrelativistic hydrogen-like atom
\begin{equation} \label{RES_rNR}
rg_{1s}^{NR}(r) = 2r\left(\frac{Z}{r_B}\right)^{3/2} e^{-r Z/r_B}.
\end{equation}

The transformed function~$r\hU f_{1S}$, which is the lower component of~$\chi_{1S}$,
should be identically zero for all~$r$.
As seen in Fig.~5, this property is satisfied with high accuracy, except in the
the vicinity of~$r=0$. To estimate the accuracy of calculations for~$r\hU f_{1S}$
we compute the norm of this function
\begin{equation} \label{RES_N}
 {\cal N}_{r\!f_{1S}} = \int_{0}^{\infty} |f_{1S}(r)| r^2 dr.
\end{equation}
Similarly, we calculate norms of~$rg_{1S}$,~$rf_{1S}$,~$r\hU f_{1S}$ and~$r\hU^{FW} f_{1S}$
for several values of the nuclear charge~$Z$. The results are listed in Table~IV.
Since the function~$r\hU f_{1S}$ should be identically zero, its norm in Eq.~(\ref{RES_N})
should vanish. In practice, the norm of~$r\hU f_{1S}$ is slightly different from zero
but negligibly small compared with the norm of~$rf_{1S}$.
This occurs both for~$Z=1$, and for~$Z=92$.

It is seen in Table~IV that the norm of~$r\hU f_{1S}$ is always smaller
than the norm of~$r\hU^{FW} f_{1S}$ which means that,
as expected, the exact transformation is more accurate than the approximate one.
On the other hand, for small~$Z$ the FW-like transformation is almost as accurate as
the Eriksen transformation. However, as seen in Table~IV, irrespective of its poorer accuracy,
the FW-like transformation for the~$1S_{1/2}$ state is quite accurate for all~$Z \in [1\ldots 92]$.

Returning to Fig.~5, it is seen that functions~$r\hU g_{1S}$ and~$r\hU^{FW} g_{1S}$ are
almost identical, i.e. both transformations lead to similar results.
Next, the transformed functions are closer to~$rg_{1S}$ function of the~$1S_{1/2}$ state
of relativistic hydrogen-like atom than to the~$1s$ state of Schrodinger hydrogen-like atom.
Since the latter function is close to~$rg_{1S}$ state of Dirac hydrogen-like atom, it is expected that
also Eriksen and FW-like transformation of the non-relativistic function would be close to~$r\hU g_{1S}$
function.

Finally, with a high accuracy,~$r\hU g_{1S}$ and~$r\hU^{FW} g_{1S}$ are normalized to unity.
Using Fig.~5 and Table~IV, one can for the first time compare results of FW-like transformation
with the results of the exact transformation separating positive and negative energy states
for the Dirac Hamiltonian with nontrivial potential.
Figure~5 and Table~IV prove high accuracy of the FW-like transformation often
used in the nonrelativistic quantum mechanics.

\begin{table}
\label{Table_Norm}
\caption{Norms of~$rg(r)$ and~$rf(r)$ functions prior to and after Eriksen and FW-like transformations,
         calculated using Eq.~(\ref{RES_N}) for different nuclear charge~$Z$.}
\begin{tabular}{|c|c|c|c|c|c|}
 \hline
  Z & Element & ${\cal N}_{r\!g}$ & ${\cal N}_{r\!f}$ & ${\cal N}_{r\hU f}$ & ${\cal N}_{r\hU^{FW}\!f}$\\
  \hline
  1  & H  & 1.000 & 1.3$\times$ 10$^{-5}$ & 1.01$\times$ 10$^{-10}$ & 1.02$\times$ 10$^{-10}$ \\
  24 & Cr & 0.992 & 0.008                 & 3.57$\times$ 10$^{-7} $ & 4.28$\times$ 10$^{-6} $ \\
  47 & Ag & 0.970 & 0.030                 & 7.62$\times$ 10$^{-6} $ & 9.51$\times$ 10$^{-5} $ \\
  74 & W  & 0.921 & 0.079                 & 5.27$\times$ 10$^{-5} $ & 7.86$\times$ 10$^{-4} $ \\
  92 & U  & 0.871 & 0.129                 & 1.33$\times$ 10$^{-4} $ & 2.29$\times$ 10$^{-3} $ \\
 \hline
\end{tabular}
\end{table}

\begin{figure}
\includegraphics[width=8.0cm,height=8.0cm]{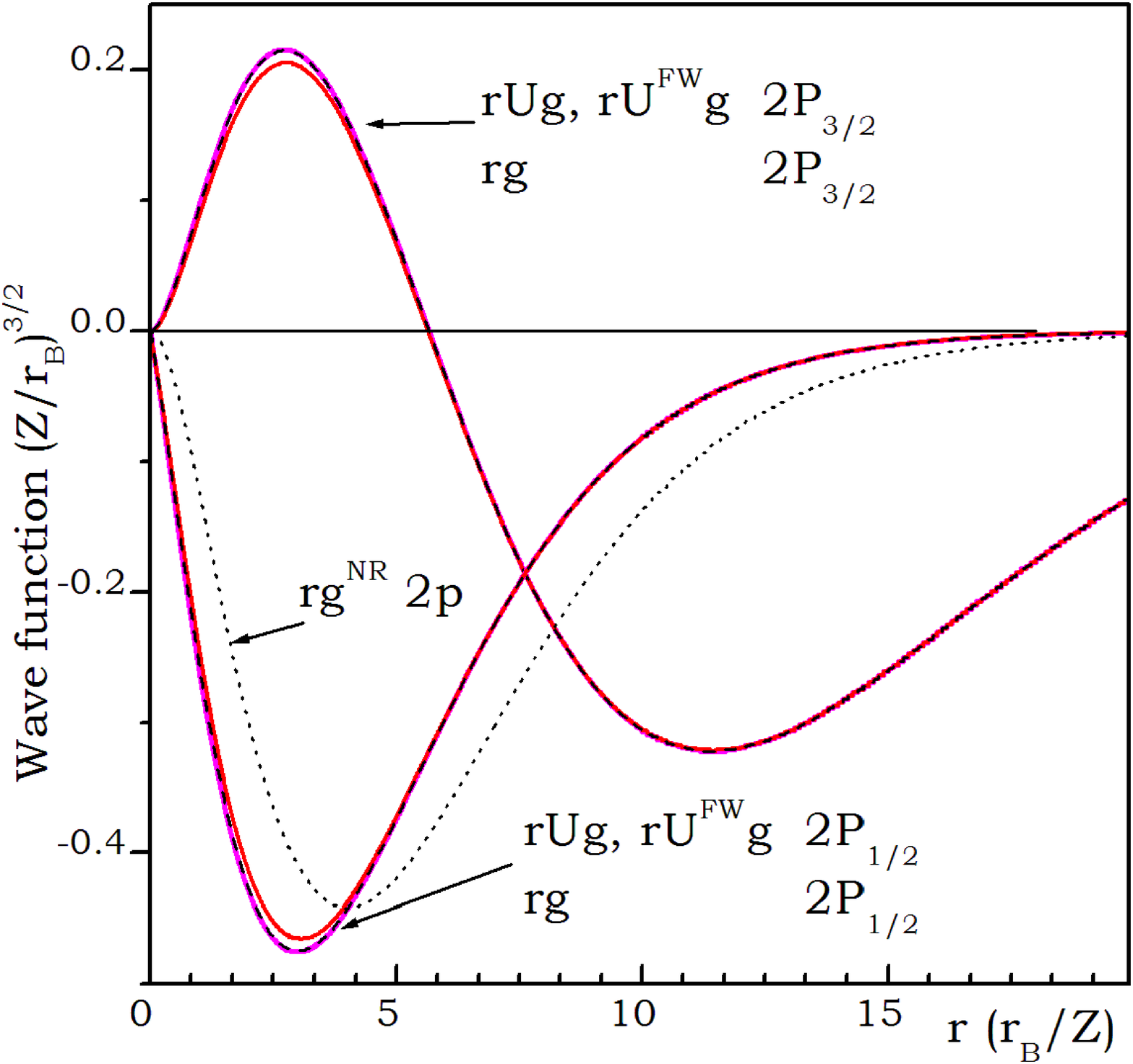}
\caption{Upper components of~$2P_{1/2}$ and~$2P_{3/2}$ states of
        relativistic hydrogen-like atom for~$Z=92$ prior to and after Eriksen and FW-like transformations.
        Note small differences between functions resulting from Eriksen and FW-like
        transformations. The functions~$rg$,~$r\hU g$ and~$r\hU^{FW} g$ for~$j=1/2$
        are close do the~$2p$ function of the nonrelativistic
        hydrogen-like atom, but for~$j=3/2$ they differ qualitatively from the~$2p$ function.} \label{Fig6}
\end{figure}

We also performed calculations of transformed functions of~$2P_{1/2}$ and~$2P_{3/2}$ states of
the relativistic hydrogen-like atom. The results for matrix elements and contributions of
various states to resulting functions are similar to those presented in Figs.~2,~3,~4 and Tables~III and~IV.
In Fig.~6 we plot the upper components of~$2P_{1/2}$ and~$2P_{3/2}$
functions prior to and after the Eriksen and FW-like transformations.
For both values of~$j$ the functions after Eriksen and FW-like transformations are
practically indistinguishable. For~$j=1/2$ and~$j=3/2$ the functions~$r\hU g$
are close to the original (non-transformed)
functions of~$2P_{1/2}$ and~$2P_{3/2}$ states of the relativistic hydrogen-like atom.
For~$j=1/2$ the function~$r\hU g$ is close to the radial function~$2p$ of the
nonrelativistic hydrogen-like atom, while the transformed function of the~$2P_{3/2}$
state is qualitatively different from the~$2p$ state.

\section{Discussion} \label{Sec_Disc}

The results presented in Sec.~\ref{Sec_Res} are obtained for the nuclear
charge~$Z \in [1 \ldots 92]$. Strictly speaking, they are valid only for
the hydrogen atom, since in real atoms the presence of many electrons and interactions between them
modify forms of atomic orbitals and continuum states, so that they cease to be the
eigenstates of~$\hH$ in Eq.~(\ref{ET_hH}). To overcome this problem we may, to some extent,
approximate the presence of other electrons by introducing an additional phenomenological
potential acting on electrons in atomic orbitals. Following Ref.~\cite{BetheBook},
for the~$1s$ state the Coulomb potential in Eq.~(\ref{ET_hH}) should be replaced by
\begin{equation} \label{Di_VZ}
V_{eff}(r)= -\frac{1}{4\pi\epsilon_0} \frac{e^2(Z-0.3)}{r} + V_0(Z),
\end{equation}
in which~$V_0(Z)$ does not depend on~$r$. After this modification, the new potential
is still Coulomb-like but with the effective value of~$Z^*=Z-0.3$.
As seen in Fig.~4, differences between the four probability densities calculated
for~$Z$ and~$Z^*$ are small. Therefore, the results shown in Figs.~2 to~5 for~$Z=92$
are very close to those computed for~$Z^*=91.7$. For~$2p$ or~$2s$ states,
the potential in Eq.~(\ref{Di_VZ}) should have a different value of the effective
charge~$Z^*<Z$~\cite{BetheBook} and the transformed~$p$-like functions in the effective
potential in Eq.~(\ref{Di_VZ}) should be close to the corresponding functions in Fig.~6.
We conclude that the results shown in Figs.~2 to~6 and Tables~III and~IV remain
valid for all~$Z \in [1 \ldots 92]$ in the range of applicability of the
approximation given in Eq.~(\ref{Di_VZ}).

It is not meaningful to analyze higher terms of the expansion of~$\hU$ in
power series of~$\alpha Z$, since one neglects the presence of radiative corrections
to the Coulomb potential which may be of comparable magnitude.
In our work we do not expand~$\hU$ in a series, but calculate~$\hU\psi$ exactly
in a finite but large basis. Within our approach we may incorporate QED effects by adding
to the Coulomb potential an additional potential~$V_{rad}(r)$.
This potential is short-range, has spherical symmetry and vanishes
for~$r\gg \lambda_c$~\cite{AkhiezerBook,Eides2001,Flambaum2005}.

As pointed out by Eriksen, a transformation analogous to that in Eq.~(\ref{ET_hU})
can be applied to two-body interactions~\cite{Eriksen1958}.
Similarly, within the Hartree-Fock approximation, the Coulomb potential
in Eq.~(\ref{ET_hH}) can be replaced by a self-consistent
potential with spherical symmetry~\cite{Nakajima2003} and our approach can be generalized
to many-electron systems.
In Sec.~\ref{Sec_Res} we limit the analysis to the lowest bound states of
relativistic hydrogen-like atom. Our method could be easily generalized to computations of
the Eriksen transformation of other bound eigenstates of~$\hH$, as described by other
quantum numbers~$\kappa$,~$m$ and~$n$.
A convenient feature of the relativistic Coulomb problem is its radial symmetry,
which limits the number of states entering into summations in Eqs.~(\ref{EC_an})--(\ref{EC_Api})
to those having the same quantum numbers~$\kappa$ and~$m$. This
simplifies computation of matrix elements in Eq.~(\ref{EC_lista}) by
reducing them to one-dimensional integrals, see Eq.~(\ref{SR_kkmm}).
This simplification is not essential to our problem,
but in practice reduces sizes of matrices~$\hS$ and~$\hZ$,
see Eqs.~(\ref{HS_Sl}) and~(\ref{HS_Zl}).

The advantage of the present approach, as compared to the series of FW-or DK-like transformations,
is a clear physical interpretation of states entering in the transformed functions~$\hU\psi$.
As seen in Figs.~3 and~4, the state~$\hU\psi$
consists mostly of the initial state~$\psi$ and continuum states having negative
energies in the vicinity of~$E\simeq -m_0c^2-E_b$. Our figures show marginal contributions
of other bound or continuum states to~$\hU\psi$.
The selection rules in Eq.~(\ref{SR_kkmm}) automatically choose proper angular
symmetry of the transformed function~$\hU\psi$. Radial functions forming~$\chi_{1S}({\bm r})$
have the same quantum numbers~$\kappa$ and~$m$ as the initial state~$\psi$.
In contrast, the Gaussian or exponential orbitals, frequently used in numerical
calculations in quantum chemistry, are somewhat artificial.

Our results shown in Figs.~3 and~4 possibly explain the accuracy of eliminating
negative-energy components in FW-like and DK-like methods, as reported in
Refs.~\cite{Leeuwen1994,Lenthe1996,Barysz2002,Nakajima2003,Reiher2004}. It is seen that
the contribution of negative-energy states to the transformed functions~$\hU\psi$ varies
from few promiles of the total probability density for light atoms
to~14\% of the total probability density for heavy atoms.
Therefore, for all nuclear charges~$Z \in [1 \ldots 92]$, the presence of negative energy
states in the function~$\hU\psi$ can always be treated as a perturbation to states of~$\hH$
having positive energies and it can be effectively removed by series of
FW-like or DK-like transformations.

The Moss-Okninski transformation~\cite{Moss1976} was used by Rusin and Zawadzki to transform a Gaussian
wave packet in the presence of an external magnetic field~\cite{Rusin2012}.
After the transformation the packet preserved the bell-like shape but its width
have changed~\cite{Rusin2012}. As seen in Figs.~5 and~6, the shape of
wave functions of the relativistic hydrogen-like atom is also retained after transformations.
The results in Fig.~5 and Table~IV for the
FW-transformed functions agree with those of Silenko~\cite{Silenko2013}
who found that the lower components of any function transformed by the FW-like
transformation are of the second order in~$1/(m_0c^2)$.

Precision of the results in Sec.~\ref{Sec_Res} depends on the accuracy
of special functions in Eqs.~(\ref{RF_rgd})--(\ref{RF_rfd})
and~(\ref{RF_rgc})--(\ref{RF_rfc}), namely the confluent hypergeometric functions,
gamma functions and hypergeometric functions. These functions have been calculated
using methods described in detail in Ref.~\cite{Pearson2014} and tested with
the results obtained on a Web-page calculator for special functions~\cite{WebCasio}.
Functions~$g(r)$ and~$f(r)$ calculated numerically were checked for
their orthogonality to other functions. The exact results for diagonal and nondiagonal
matrix elements of~$\hB$ and~$\hS$ served as an additional tests of the employed procedures.
Asymptotic forms of~$g(r)$ and~$f(r)$ were used for testing
exact functions~$g(r)$ and~$f(r)$ in Eqs.~(\ref{RF_rgc})--(\ref{RF_rfc})
and their normalization. A convenient feature of our problem is a
possibility of analytical calculations for all matrix
elements of~$\hB$ in Eq.~(\ref{EC_lista}).
There are two tests of accuracy of the numerical procedures:
the sum-rule for the probability amplitudes~$a_n$ and~$A_{p_i}^{\epsilon}$ in Eq.~(\ref{EC_sumAa}),
and the requirement that function~$\hU\psi$ has vanishing lower components.
As seen in Fig.~5 and Table~IV, a high accuracy of numerical calculation is achieved.

\section{Summary} \label{Sec_Summ}

We calculated the single-step Eriksen transformation of the wave
functions for~$1S_{1/2}$,~$2P_{1/2}$ and~$2P_{3/2}$ states of the relativistic hydrogen-like
atom. In the new representation the functions have two nonzero components.
The proposed method does not require an expansion of the Eriksen operator~$\hU$
in the power series of~$1/(m_0c^2)$ or the potential. Our approach is based
on the observation that, although the operator~$\hU$ defining the transformation
is not given in an explicit form, it is possible to calculate analytically
or numerically its matrix elements between eigenstates of~$\hH$.
To exploit this observation, we expressed the transformed wave functions in
the form of linear combinations of eigenstates of~$\hH$
for a sufficiently large set of states. The continuum states of~$\hH$
are replaced by the so-called discretized functions (eigendifferentials),
which allow one to treat bound and continuum states in a similar way.
Our results may possibly explain the accuracy of FW-like and
DK-like transformations reported in literature, since the contribution from states of
negative energies to the total probability density can be safely
treated as a perturbation to the contribution of states with positive energies.
As expected, lower components of~$\hU\psi$ are nearly zero except in the
vicinity of~$r=0$. This result confirms accuracy of the Eriksen transformation
and numerical calculations. Upper components of~$\hU\psi$ are
well-localized functions, similar to their counterparts in the
Dirac representation. The non-vanishing components
of~$\hU\psi$ and~$\hU^{FW}\psi$ are close to each other
which confirms the accuracy of the FW-like transformation.
Concluding, it is believed that the
reported results contribute to better understanding of the Eriksen and
Foldy-Wouthuysen transformations.

\acknowledgements
I acknowledge Mrs. Aneta Osowska and Prof. Wlodek Zawadzki for help and discussions during the
preparation and reading of the manuscript. I acknowledge Casio Company for designing and operating
the Free On-Line calculator for the special functions, see Ref.~\cite{WebCasio}, which was used for
validation of the numerical procedures calculating the radial functions of the
relativistic hydrogen-like atom.

\appendix

\section{Properties of~\mbox{$\hZ$} and~\mbox{$\hU$} operators} \label{Appendix_Z}

Here we show that the operator~$\hZ$ in Eq.~(\ref{ET_hZ}) has vanishing matrix elements between
eigenstates of~$\hH$ for positive and negative energies. Consider matrix~$\hS$ in Fig.~1, which is block-diagonal
with vanishing matrix elements between eigenstates of~$\hH$ having positive and negative
energies. Schematically, in the basis used in Fig.~1, its form
is~$\hS=\left(\begin{array}{cc}\hS^+&0\\ 0 &\hS^- \end{array} \right)$,
where operators~$\hS^{\pm}$ are constructed from matrix elements of~$\hB$ between states having
the same energy signs. Then, any integer power of~$\hS$ is also
block-diagonal:~$(\hS)^m=\left( \begin{array}{cc}(\hS^+)^m&0\\ 0 & (\hS^-)^m \end{array} \right)$.
The operator~$\hZ$ can be expanded in power series:~$\hZ = \sum_{m=0}^{\infty} z_m (\hS)^m$,
with suitably chosen coefficients~$z_m$, see Ref.~\cite{Eriksen1958}.
Since each term of this series is block-diagonal, the operator~$\hZ$ is block-diagonal as well,
and its matrix elements between eigenstates of~$\hH$ for different energy signs vanish.
This completes the proof.

The operator~$\hZ$ has block-diagonal form in representation of the Dirac spinors.
To show this we first note that~$\hS$ is block-diagonal in this representation.
Expressing~$\hl$ in form of~$2\times 2$
blocks:~$\hl=\left(\begin{array}{cc}\hl_{11}& \hl_{12}\\ \hl_{21} &\hl_{22} \end{array} \right)$
there is:~$\hS=\hB\hl+\hl\hB = \left(\begin{array}{cc}2\hl_{11}& 0 \\ 0 &-2\hl_{22} \end{array} \right)$.
Repeating the arguments presented above we
find:~$\hZ= \left(\begin{array}{cc}\hZ_{11}& 0 \\ 0 &\hZ_{22} \end{array} \right)$.

Next we show that~$\hU$ in Eq.~(\ref{ET_hU}) transforms any eigenstate of the Dirac
Hamiltonian to the two-component form.
Let~$\psi^{\pm}=\left(\begin{array}{cc}g^{\pm}\\ f^{\pm} \end{array} \right)$
be eigenstates of~$\hH$ of positive or negative energies, respectively, and~$g^{\pm}$,~$f^{\pm}$
be the two-component vectors. Then~$\hl\psi^{\pm}= \pm\psi^{\pm}$ and
\begin{eqnarray} \label{AppZEp}
\hU \psi^+ &=& \hZ\left (1 +\hB\hl\right)\psi^+ =
 \left(\begin{array}{cc} \hZ_{11} g^+\\ 0\end{array} \right), \\
 \label{AppZEm}
\hU \psi^-& =& \hZ\left(1 +\hB\hl\right)\psi^- =
 \left(\begin{array}{cc} 0\\ \hZ_{22} f^-\end{array} \right).
\end{eqnarray}
This completes the proof.

The operators~$\hB\hl$ and~$\hl\hB$ commute, which can be shown directly.
Expanding~$\hZ$ in a power series of~$(\hS-2)$ we find that each term of the series
commutes with~$\hB\hl$ and~$\hl\hB$. Therefore~$\hZ$ commutes with~$\hB\hl$ and~$\hl\hB$,
which allows one to interchange the orders of~$\hZ$ and~$\hB\hl$
in the definition of~$\hU$ in Eq.~(\ref{ET_hU}).

\section{Matrix elements of~\mbox{$\hB$}} \label{Appendix_Beta}

The diagonal matrix elements of~$\hB$ can be obtained without
direct specification of the wave functions~$\langle {\bm r}|{\rm n} \rangle$, see e.g. Ref.~\cite{Goldman1982}.
Here we calculate these elements with the use of the Hellmann-Feynman theorem.
Let us treat the electron mass~$m_0$ in Eq.~(\ref{ET_hH}) as a variable parameter. Then
\begin{equation} \label{AppMEB_0}
\langle {\rm n}(m_0)|\hH(m_0)|{\rm n}(m_0)\rangle = E(m_0).
\end{equation}
Differentiating both sides of Eq.~(\ref{AppMEB_0}) with respect of~$m_0$ one obtains
\begin{equation} \label{AppMEB_Bnn}
\langle {\rm n}|\hB|{\rm n}\rangle = \frac{1}{c^2} \frac{\partial E(m_0)}{\partial m_0}.
\end{equation}
and the diagonal matrix elements of~$\hB$ can be obtained by straightforward
differentiations of both sides of Eqs.~(\ref{RF_En}) and~(\ref{RF_Ep}) with respect of~$m_0$.
The result of Eq.~(\ref{RF_En}) are known~\cite{Goldman1982} but, to our knowledge,
their extension to the continuum states in Eq.~(\ref{RF_Ep}) has not been published yet.

Next, we consider the diagonal matrix elements between discretized functions defined in Eq.~(\ref{DISC_1}).
Combining Eqs.~(\ref{DISC_1}),~(\ref{RF_Ep}) and~(\ref{AppMEB_Bnn}) one has
\begin{eqnarray}
\langle \Psi_{p_i}^{\epsilon_i}|\hB|\Psi_{p_j}^{\epsilon_j} \rangle=
                \frac{1}{\Delta p} \int_{p_i-\Delta p/2}^{p_i+\Delta p/2}
                                   \int_{p_j-\Delta p/2}^{p_j+\Delta p/2}
                \langle \psi_{p_1}^{\epsilon_i}|\hB| \psi_{p_1}^{\epsilon_j}\rangle dp_1dp_2 \nonumber \\
 = \frac{\delta_{\epsilon_i, \epsilon_j}}{\Delta p} \int_{p_i-\Delta p/2}^{p_i+\Delta p/2}
                                   \int_{p_j-\Delta p/2}^{p_j+\Delta p/2}
                                   \frac{\partial E_{p_1}}{c^2\partial m_0} \delta(p_1-p_2) dp_1dp_2 \nonumber \\
\label{AppMEB_B}
 = \frac{\delta_{\epsilon_i, \epsilon_j}\delta_{p_i,p_j}}{\Delta p} \int_{p_i-\Delta p/2}^{p_i+\Delta p/2} \frac{m_0c^2}{E_p} dp
 \ \ \ \ \ \ \ \ \ \ \ \ \ \\
\label{AppMEB_Bapr}
 \xrightarrow[\Delta p \rightarrow 0]\
 \delta_{\epsilon_i, \epsilon_j}\delta_{p_i,p_j} \frac{m_0c^2}{E_{p_i}}. \ \ \ \ \ \ \ \ \ \ \ \ \
\end{eqnarray}
The energy~$E_p$ in Eq.~(\ref{AppMEB_B}) is given in Eq.~(\ref{RF_Ep}).
Using the same approach to the diagonal matrix elements of~$\hZ$ between
discretized states we find
\begin{eqnarray} \label{AppMEB_Z}
\langle \Psi_{p_i}^{-}|\hZ|\Psi_{p_j}^{-} \rangle &=&
 \frac{\delta_{p_i,p_j}}{\Delta p} \int_{p_i-\Delta p/2}^{p_i+\Delta p/2} \frac{dp}{\sqrt{2+2m_0c^2/E_p}} \nonumber \\
 &\simeq & \frac{\delta_{p_i,p_j}}{\sqrt{2+2m_0c^2/E_{p_i}}}.
\end{eqnarray}
Since functions~$\psi_p^{\epsilon}$ are normalized to~$\delta(p_1-p_2)\delta_{\epsilon_1\epsilon_2}$,
the nondiagonal matrix elements of~$\hB$ and~$\hZ$ between continuum functions vanish.

A calculation of the matrix element of~$\hB$ between discretized states in Eq.~(\ref{DISC_2})
and the bound ones requires computation of the integral
\begin{eqnarray} \label{AppMEB_Psi}
 \langle \Psi_{p_i}^{\epsilon}|\hB|\text{1S} \rangle
 &=& \frac{1}{\sqrt{\Delta p}} \int_{p_i-\Delta/2}^{p_i+\Delta/2} \langle \psi_p^{\epsilon}|\hB|\text{1S} \rangle dp \nonumber \\
 &\xrightarrow [\Delta p \rightarrow 0]& \sqrt{\Delta p}\ \langle \psi_{p_i}^{\epsilon}|\hB|\text{1S} \rangle.
\end{eqnarray}

The element~$\hB_{1S,p-}$ is a combination of two integrals of the form
\begin{equation} \label{AppMEB_Q1}
 Q^{\epsilon} = T_0\int_0^{\infty} e^{-(C_0+i\epsilon K)r}(2rC_0)^{\gamma}(2Kr)^{\gamma} {_1}F_1(a,\mu,2iKr) dr,
\end{equation}
in which~$T_0$ an is~$r$-independent constant [see Eqs.~(\ref{RF_rgd})--(\ref{RF_rfd})
and~(\ref{RF_rgc})--(\ref{RF_rfc})],~$a=\gamma+1+iy$,~$\mu= 2\gamma+1$,~$C_0=Z/r_B$,
and~$K=(p/\hbar)(\alpha Z)$. The integral in Eq.~(\ref{AppMEB_Q1}) can be calculated analytically and we obtain
\begin{equation} \label{AppMEB_Q2}
 Q^{\epsilon}= T_0\Gamma(\mu)(4C_0)^{\gamma}\frac{K^{\gamma}}{(C_0+i\epsilon K)^{\mu}}
 \left(\frac{C_0-i\epsilon K}{C_0+i\epsilon K}\right)^{a}.
\end{equation}
The remaining matrix elements~$\hB_{nS,p\epsilon}$ can be expressed in terms of special functions and below we list formulas
necessary for the calculations. For bound states the confluent hypergeometric function is
the polynomial defined recurrently
\begin{eqnarray} \label{AppMEB_CHG}
{_1}F_1\left(-n,c,z \right) &=& \sum_{j=0}^{n} t_{n,j}\ z^j, \\
\label{AppLS_CHG2}
 t_{n,j+1} &=& t_{n,j}\frac{-n+j-1}{j(c+j-1)},
\end{eqnarray}
with~$t_{n,0}=1$.
The matrix elements~$\hB_{nS,mS}$ between bound functions can be calculated with use of the Euler integral
\begin{equation} \label{AppMEB_G}
\int_0^{\infty} x^{b-1}e^{-s x} dx = \frac{1}{s^b} \Gamma(b).
\end{equation}
For calculations of the matrix elements~$\hB_{nS,p\epsilon}$ between bound
and continuum states we use the formula~\cite{WebDLMF}
\begin{equation}
\label{AppB_Lap}
 \int_0^{\infty} e^{-st}t^{b-1}{_1}F_1(a,c,kt)dt = \frac{\Gamma(s)}{z^{b}} {_2}F_1(a,b,c,k/s),
\end{equation}
assuming that~$\Re(b)>0$ and~$\Re(s)>\max(\Re(k),0)$, where~${_2}F_1(a,b,c,z)$ is the hypergeometric function
in the standard notation~\cite{GradshteinBook}. We use also the
identities:~${_2}F_1(a,b,c,z) = {_2}F_1(b,a,c,z)$ and~${_2}F_1(a,b,c,z) = (1-z)^{-a} {_2}F_1(a,c-b,c,z/(z-1))$.
More general expressions for the matrix elements of~$\hB$ are given in Ref.~\cite{Suslov2009,Reynolds1964,Gargaro1970}.


\begin{thebibliography}{99}
\bibitem{Foldy1950}       L. L. Foldy and S. A. Wouthuysen, Phys. Rev. {\bf 78}, 29 (1950).
\bibitem{deVries1970}     E. de Vries, Fortschritte der Physik {\bf 18}, 149 (1970).
\bibitem{Douglas1974}     M. Douglas and N. M. Kroll, Ann. Phys. (N. Y.) {\bf 82}, 89 (1974).
\bibitem{Jansen1989}      G. Jansen and B. A. Hess, Phys. Rev. A {\bf 39}, 6016 (1989).
\bibitem{Nakajima2003}    T. Nakajima and K. Hirao, J. Chem. Phys. {\bf 119}, 4105 (2003).
\bibitem{Reiher2015}      M. Reiher, arXiv: 1501.05764 v2 (2015).
\bibitem{Silenko2003}     A. J. Silenko, J. Math. Phys. {\bf 44}, 2952 (2003).
\bibitem{Reiher2004}      M. Reiher and A. Wolf, J. Chem. Phys. {\bf 121}, 10945 (2004).
\bibitem{Leeuwen1994}     R. van Leeuwen, E. van Lenthe, E. J. Baerends, and J. G. Snijders, J. Chem. Phys. {\bf 101}, 1272 (1994).
\bibitem{Lenthe1996}      E. van Lenthe, E. J. Baerends, and J. G. Snijders, J. Chem. Phys. {\bf 105}, 2373 (1996).
\bibitem{Barysz2002}      M. Barysz and A. J. Sadlej, J. Chem. Phys. {\bf 116}, 2696 (2002).
\bibitem{Case1954}        K. M. Case, Phys. Rev. {\bf 95}, 1323 (1954).
\bibitem{Tsai1973}        W. Y. Tsai, Phys. Rev. D {\bf 7}, 1945 (1973).
\bibitem{Weaver1975}      D. L. Weaver, Phys. Rev. D {\bf 12}, 4001 (1975).
\bibitem{Moss1976}        R. E. Moss and A. Okninski, Phys. Rev. D {\bf 14}, 3358 (1976).
\bibitem{Nikitin1998}     A. G. Nikitin, J. Phys. A: Math. Gen. {\bf 31}, 3297 (1998).
\bibitem{Rusin2011}       T. M. Rusin and W. Zawadzki, Phys. Rev. A {\bf 84}, 062124 (2011).
\bibitem{Rusin2012}       T. M. Rusin and W. Zawadzki, J. Phys. A: Math. Theor. {\bf 45}, 315301 (2012).
\bibitem{Neznamov2009}    V. P. Neznamov and A. J. Silenko, J. Math. Phys. {\bf 12}, 12230 (2009).
\bibitem{Silenko2008}     A. J. Silenko, Physics of Particles and Nuclear Letters {\bf 5}, 501 (2008).
\bibitem{Eriksen1958}     E. Eriksen, Phys. Rev. {\bf 111}, 1011 (1958).
\bibitem{Eriksen1960}     E. Eriksen and M. Kolsrud, Nuovo Cimento Suppl. {\bf 18}, 1 (1960).
\bibitem{Silenko2013}     A. J. Silenko, Physics of Particles and Nuclear Letters {\bf 10}, 198 (2013).
\bibitem{BransdenBook}    B. H. Bransden and M. R. C. McDowell {\it Charge Exchange and the Theory of Ion-Atom Collisions}
                          (Clarendon Press, Oxford, 1992).
\bibitem{Kocbach2001}     L. Kocbach and I. Ladadwa, AIP Conf. Proc. {\bf 576}, 68 (2001).
\bibitem{Bertulani1992}   C. A. Bertulani and F. Canto, Nucl. Phys. A {\bf 539}, 163 (1992).
\bibitem{GreinerIntro}    W. Greiner {\it Quantum Mechanics: An Introduction} (Springer, Berlin, 2001).
\bibitem{RoseBook}        M. E. Rose {\it Relativistic Electron Theory} (Wiley, New York, 1961).
\bibitem{GreinerBook}     W. Greiner {\it Relativistic Quantum Mechanics} (Springer, Berlin, 1994).
\bibitem{AkhiezerBook}    A. I. Akhiezer and V. B. Berestetskii {\it Quantum Electrodynamics} (Wiley, New York, 1965).
\bibitem{GradshteinBook}  I. S. Gradshtein and I. M. Ryzhik in {\it Table of Integrals, Series, and Products},
                          7th ed., edited by A. Jeffrey and D. Zwillinger (Academic Press, New York, 2007).
\bibitem{Rose1937}        M. E. Rose, Phys. Rev. {\bf 51}, 484 (1937).
\bibitem{Goldman1982}     S. P. Goldman and G. W. F. Drake, Phys. Rev. A {\bf 25}, 2877 (1982).
\bibitem{Suslov2009}      S. K. Suslov, J. Phys. B: Atomic Mol. Opt. Phys. {\bf 42}, 185003 (2009).
\bibitem{Reynolds1964}    J. T. Reynolds, D. S. Onley, and L. C. Biedenharn, J. Math. Phys. {\bf 5}, 411 (1964).
\bibitem{Gargaro1970}     W. W. Gargaro and D. S. Onley, J. Math. Phys. {\bf 11}, 1191 (1970).
\bibitem{BetheBook}       H. A. Bethe and E. E. Salpeter {\it Quantum Mechanics of One- and Two-electron Atoms}
                          (Academic Press, New York, 1957).
\bibitem{Eides2001}       M. I. Eides, H. Grotch, and V. A. Shelyuto, Phys. Rep. {\bf 342}, 63 (2001).
\bibitem{Flambaum2005}    V. V. Flambaum and J. S. M. Ginges, Phys. Rev. A {\bf 72}, 052115 (2005).
\bibitem{Pearson2014}     J. W. Pearson, S. Olver, and M. A. Porter, arXiv:1407.7786v2 (2014).
\bibitem{WebCasio}        http://keisan.casio.com/menu/system/000000000760, (2016).
\bibitem{WebDLMF}         http://dlmf.nist.gov/13.10, Eq. 13.10.3, (2016).
\end{thebibliography}
\end{document}